\documentstyle[11pt,paspconf,epsf]{article}
\begin{document}

\title{A Survey for Low Surface Brightness Dwarf Galaxies Around M31}
\author{Taft E.\ Armandroff}
\affil{Kitt Peak National Observatory, National Optical Astronomy
Observatories,\altaffilmark{1} P.O.\ Box 26732, Tucson, AZ 85726}
\author{James E.\ Davies\altaffilmark{2}}
\affil{Department of Astronomy, University of Wisconsin, Madison, WI
53706}
\author{George H.\ Jacoby}
\affil{Kitt Peak National Observatory, National Optical Astronomy
Observatories,\altaffilmark{1} P.O.\ Box 26732, Tucson, AZ 85726}
\altaffiltext{1}{The National Optical Astronomy Observatories are
operated by the Association of Universities for Research in
Astronomy, Inc., under cooperative agreement with the National
Science Foundation.}
\altaffiltext{2}{Currently at Dept.\ of Physics and Astronomy,
Johns Hopkins University, Baltimore, MD 21218.}

\begin{abstract}
By applying a digital filtering technique to 1550 deg$^2$ of the
POSS-II in the vicinity of M31, we found two previously unidentified
very low surface brightness dwarf galaxies which we designate And V and
VI.\@ Follow-up imaging with the KPNO 4-m telescope resolved these into
stars easily. The $V$- and $I$- band images of And V indicate a
distance similar to that of M31, and $<$[Fe/H]$>$ $\sim$ --1.5.\@ All
evidence strongly supports its classification as a dwarf spheroidal
companion to M31.\@ Data for And VI are being analyzed, but preliminary
indications support a similar conclusion.  Our search for more dwarfs,
including follow-up observations of numerous candidates found via
digital filtering, is incomplete; thus, further identifications may be
forthcoming.
\end{abstract}


\section{Introduction}
In order to understand many aspects of galaxy formation and galaxy
evolution, a complete and unbiased census of the members of the Local
Group is necessary.  Some examples of the problems that can be
addressed with such knowledge include:  1) Accurately defining the
faint end of the galaxy luminosity function for the Local Group,
allowing comparison with that in rich clusters of galaxies, for
example.  2) Identifying additional dynamical probes within the Local
Group in order to facilitate deciphering its dynamical history and
total mass.

A survey for low surface brightness dwarf galaxies in the direction of
M31 yields additional benefits.  With a complete census of M31's dwarf
companions and knowledge of their properties, we can compare M31's
satellite system with that of the Galaxy.  Any differences or
similarities between the two systems yield information on the effects
of environment on the formation and evolution of dwarf galaxies.  In
addition, the properties of M31's dwarf companions tell us how much we
can generalize from the Galaxy's dwarf satellite system.

The study of M31's dwarf spheroidal companions began with van den
Bergh's (1972a, 1974) search of $\sim$700 square degrees around M31
using IIIaJ photographic plates taken with the Palomar 48-inch Schmidt
telescope.  By visually examining the plates, he identified three dwarf
spheroidal candidates (And I, II \& III).\@  By taking deeper plates
with the Palomar 5-m telescope, van den Bergh (1972b, 1974) found that
And I, II \& III resolve into stars at approximately the same magnitude
as M31's companion NGC 185, strengthening the association of these
galaxies with M31.\@  Recent color--magnitude diagrams for And I, II \&
III from large terrestrial telescopes (Mould \& Kristian 1990;
Armandroff et al.\ 1993; Konig et al.\ 1993) and Hubble Space Telescope
(Da Costa et al.\ 1996) have yielded distances that place And I, II \&
III in the outer halo of M31.\@  These deep color--magnitude diagrams
have also resulted in mean metallicity, metallicity dispersion, and age
information.  Surface photometry and structural parameters are
available for And I, II \& III from Caldwell et al.\ (1992).  All
available data suggest that And I, II \& III closely resemble the
Galactic dwarf spheroidals.

A number of arguments suggest that a new search of M31's environs for
low surface brightness dwarfs may be worthwhile.  First, the absolute
magnitudes of And I, II \& III do not extend as faint as the Galactic
dwarf spheroidals (see Figure 2 of Armandroff 1994), suggesting that
perhaps fainter dwarfs exist below the detection limit.  Second, the
area surveyed by van den Bergh (1972a, 1974) does not reach the
projected distance from the center of M31 that one would expect based
on the most distant Galactic dwarf companions (see Figure 6 of
Armandroff 1994).\@  Finally, why does M31 have three known dwarf
spheroidal companions when the Galaxy has nine?  Is this the result of
incompleteness in the case of M31 or some intrinsic difference between
the two systems?

Consequently, we have undertaken a new search for low surface
brightness dwarf galaxies around M31.\@  Our search uses digital
filtering techniques applied to the Second Palomar Sky Survey
(POSS-II).\@  Our search is complementary to the nearby dwarf galaxy
survey by Irwin (1994) that used star counts over much of the sky and
revealed the Sextans dwarf spheroidal.  Because dwarf galaxies at the
M31 distance do not resolve into stars on survey plates, our search is
not based on star counts.  Since M31 is a northern object, our search
is also complementary to that of Whiting et al.\ (1997) based on visual
examination of the southern sky survey plates that called attention to
the Antlia dwarf.

\section{Search Strategy}
The POSS-II (Reid et al.\ 1991) has higher resolution and extends
substantially deeper than its first generation counterpart.  The
availability of the POSS-II in digital form (Lasker \& Postman 1993)
enables the use of digital processing techniques and allows full areal
coverage around M31.\@  Our survey program employs the POSS-II and seeks
to: 1) search an area around M31 that is commensurate with the radial
distribution of dwarf spheroidals around the Galaxy; 2) use digital
processing techniques to enhance the detection of the faintest, lowest
surface brightness dwarf spheroidals.

Our search methodology employs a matched filter and is described fully
in Armandroff, Davies \& Jacoby (1998).\@  Briefly, the procedure
includes:  dividing each plate into overlapping 1 square degree
regions; fitting a surface to remove the background; removing stars by
clipping  values $>$ 0.75$\sigma$ above the median; applying a spatial
median filter (77 $\times$ 77 arcsec; based on And II \& III); visual
examination of the processed and original images; selection of
candidates that resemble And II \& III on these images.  Obvious
bright-star ghosts and ``optical cirrus'' clouds are avoided in the
selection process.  We check the coordinates against SIMBAD and NED to
eliminate known objects.  To date, we have applied our detection
procedure to 1550 square degrees of the POSS-II around M31.

As the next step in our survey, all the remaining candidates are imaged
with the KPNO 0.9-m telescope.  These CCD images eliminate any
candidates that are plate-based false detections (e.g., remaining
bright-star ghosts, unrecognized emulsion problems) and most
astronomical misidentifications (e.g., distant galaxy clusters,
background low surface brightness spirals, ``optical cirrus" clouds).\@
If a candidate shows incipient resolution into stars on the 0.9-m CCD
images, then it is probably a nearby dwarf galaxy, so imaging with a
4-m-class telescope is undertaken.  At the time that this paper was
written (October 1998), we had not yet surveyed all the POSS-II plates
in our desired search area, nor had we completed CCD imaging of our
list of candidates from the POSS-II.
\begin{figure}[t]
\caption{The left panels show images of And V \& VI from the
digitized POSS-II; the right panels show the results of applying our
digital enhancement procedure.  Each panel is 8.5 arcmin on a side.}
\label{and5and6dss}
\end{figure}

\section{Andromeda V}
Thus far, two candidates found by our survey have been resolved into
stars (see Figure \ref{and5and6dss} for POSS-II images of these
candidates).\@ The first of these, called Andromeda V, is discussed
extensively in Armandroff et al.\ (1998).\@ We summarize our most
important And V results below.

And V is located at the following coordinates: $\alpha$ =
1$^{\rm h}$ 10$^{\rm m}$ 17.1$^{\rm s}$, $\delta$ = +47$^\circ$
37$^\prime$ 41$^{\prime\prime}$ (J2000.0).\@
And V was observed with the KPNO 4-m telescope and prime-focus CCD
imager in the $V$ and $I$ filters, plus in H$\alpha$ narrow-band and
$R$.\@  In the broad-band filters, And V resolves nicely into stars and
exhibits a smooth stellar distribution (Figure
\ref{and5ccd}).\@  In these images, And V resembles the other M31 dwarf
spheroidals.  And V does not exhibit the features of classical dwarf
irregulars, such as obvious regions of star formation or substantial
asymmetries in its stellar distribution.  In the And V
continuum-subtracted H$\alpha$ image, no diffuse H$\alpha$ emission or
H {\sc ii} regions are detected.  The lack of H$\alpha$ emission in And
V reinforces the conclusion, based on And V's appearance on the
broad-band images, that it is a dwarf spheroidal galaxy rather than a
dwarf irregular.
\begin{figure}[t]
\caption{A $V$-band image of And V (three 900-second exposures) taken
with the KPNO 4-m telescope.  North is at the top, and east is to the
right.  The image covers $3\times3$ arcmin.}
\label{and5ccd}
\end{figure}

And V is not detected in any of the IRAS far-infrared bands either.
Because far-infrared emission traces warm dust, and because some Local
Group dwarf irregular galaxies are detected by IRAS, And V's lack of
far-infrared emission serves as additional, weaker evidence that it is
a dwarf spheroidal.  No information is currently available about the H
{\sc i} content of And V via 21 cm observations.  Either an H {\sc i}
detection or a strict upper limit would be valuable.

We have measured And V's apparent central surface brightness via
large-aperture photometry in its core: 25.7 mag/arcsec$^2$ in $V$.\@  And
V has a fainter apparent central surface brightness than And I, II \&
III (24.9, 24.8, and 25.3 mag/arcsec$^2$ in $V$, respectively;
Caldwell et al.\ 1992).\@  And V probably eluded detection until now due
to its very dim apparent surface brightness.

\begin{figure}[t]
\plotfiddle{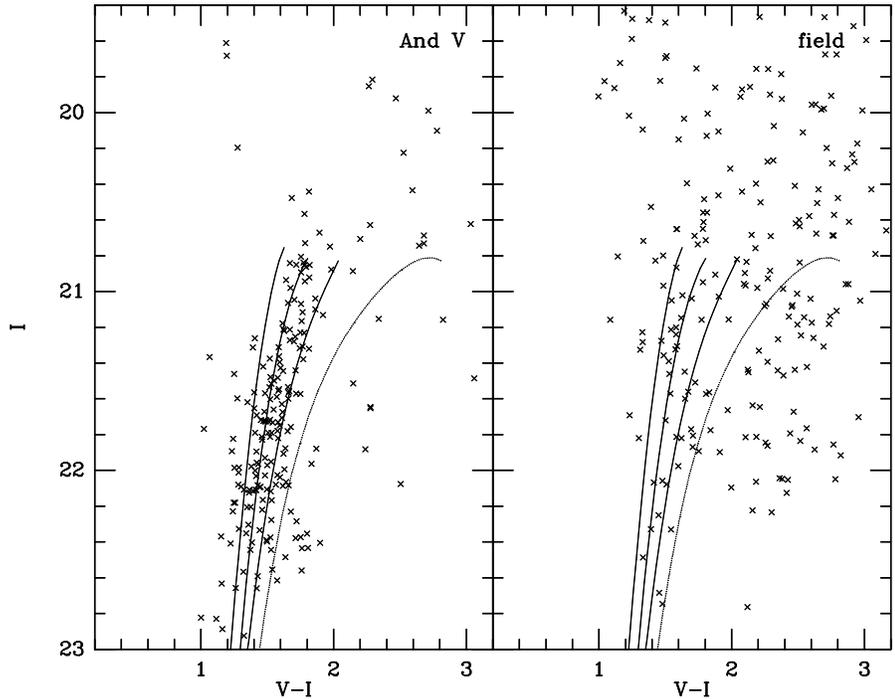}{3.6in}{270.}{50.}{50.}{-182}{283}
\caption{Color--magnitude diagrams.  The left panel shows stars within
a radius of 71 arcsec of the center of And V, where And V members
outnumber field contamination.  The right panel displays stars greater
than 168 arcsec from the center of And V, where the contribution from
And V stars is negligible.  The right panel represents 5.5 times more
area on the CCD than the left panel.  Red giant branch fiducials for
four Galactic globular clusters that span a range of metal abundance
(Da Costa \& Armandroff 1990), shifted to the distance modulus and
reddening of And V, have been overplotted.  From left to right, the red
giant branch fiducials are M15 ([Fe/H] = --2.17), M2 (--1.58), NGC 1851
(--1.16), and 47 Tuc (--0.71).}
\label{cmd}
\end{figure}

We used our $V$ and $I$ images and photometric standard observations to
construct a color--magnitude diagram for And V stars, in order to
reveal And V's distance and stellar populations characteristics.
Color--magnitude diagrams for the parts of the images dominated by And
V stars reveal a red giant branch, which is absent in the outer regions
of the images (see Figure \ref{cmd}).\@  The tip of the red giant branch
is well defined in the color--magnitude diagram and in the luminosity
function.  A distance has been derived for And V based on the $I$
magnitude of the tip of the red giant branch (Da Costa \& Armandroff
1990, Lee et al.\ 1993).\@ On the distance scale of Lee et al.\ (1990),
the resulting And V distance is 810 $\pm$ 45 kpc.\@  How does this And
V distance compare with that of M31?  The most directly comparable
distance determinations for M31, based on either red giant branch tip
stars or RR Lyraes in the M31 halo or horizontal branch stars in M31
globular clusters, are 760 $\pm$ 45 kpc or 850 $\pm$ 20 kpc, both on
the same scale (see Da Costa et al.\ 1996).\@ Our distance for And V of
810 $\pm$ 45 kpc implies that And V is located at the same distance
along the line of sight as M31 to within the uncertainties.  And V's
projected distance from the center of M31 is 112 kpc; And I, II \& III
have projected M31-centric distances of 46, 144 and 69 kpc,
respectively.  The above line-of-sight and projected distances strongly
suggest that And V is indeed associated with M31.

In order to investigate the stellar populations in And V, we have
compared its color--magnitude diagram with fiducials representing the
red giant branches of Galactic globular clusters that span a range of
metal abundance (Da Costa \& Armandroff 1990; see Figure \ref{cmd}).\@
Based on the position of the And V giant branch relative to the
globular cluster fiducials, the mean metal abundance of And V is
approximately --1.5.\@  This metallicity is normal for a dwarf
spheroidal (e.g., see Figure 9 of Armandroff et al.\ 1993).\@  No
bright blue stars are present in the And V color--magnitude diagram.
Interpreting via isochrones, this lack of blue stars rules out any
stars younger than 200 Myr in And V and is further evidence that And V
is a dwarf spheroidal and not a dwarf irregular.

From the luminosities and numbers of upper asymptotic giant branch
stars in a metal-poor stellar system, one can infer the age and
strength of its intermediate age component (Renzini \& Buzzoni 1986).\@
Using the And V field-subtracted $I$ luminosity function, there is no
evidence for upper asymptotic giant branch stars that are more luminous
than and redward of the red giant branch tip.  Therefore, And V does
not have a prominent intermediate age population; in this sense, it is
similar to And I \& III.

\section{Andromeda VI}
Using our search methodology, we found a second candidate dwarf
spheroidal galaxy, designated Andromeda VI, at the following celestial
coordinates: $\alpha$ = 23$^{\rm h}$ 51$^{\rm m}$ 46.9$^{\rm s}$,
$\delta$ = +24$^\circ$ 34$^\prime$ 45$^{\prime\prime}$ (J2000.0).\@ And
VI was imaged with the KPNO 4-m telescope prime-focus CCD on 1998
January 23 in $V$ for 300 seconds.  And VI resolved nicely into stars
in this short $V$ image, suggesting that it is indeed a nearby dwarf
galaxy.  Several months after we had confirmed to ourselves that And VI
is a bona fide nearby dwarf galaxy, we learned that Karachentsev \&
Karachentseva were reporting the discovery of this same galaxy, calling
it the Pegasus Dwarf (see Brinks \& Grebel 1998).

\begin{figure}[t]
\caption{An $R$-band image of And VI (three 300-second exposures) taken
with the KPNO 4-m telescope.  North is to the left, and east is at the
bottom.  The image covers $3.9\times3.7$ arcmin.}
\label{and6ccd}
\end{figure}
On 1998 July 15, And VI was imaged more deeply with the KPNO 4-m
telescope through the $R$ and H$\alpha$ filters.  The $R$
image of And VI, displayed in Figure \ref{and6ccd}, exhibits a smooth
stellar distribution and a resemblance with the other M31 dwarf
spheroidals.  And VI does not look lumpy or show obvious regions of
star formation that would suggest a dwarf irregular, as opposed to
dwarf spheroidal, classification.  In the And VI continuum-subtracted
H$\alpha$ image, no diffuse H$\alpha$ emission or H {\sc ii} regions
are detected.  The lack of H$\alpha$ emission rules out current
high-mass star formation in And VI and serves as further evidence that
And VI is a dwarf spheroidal.  Like And V and the other M31 dwarf
spheroidals, And VI is not detected in any of the IRAS far-infrared
bands.

We are currently working with images of And VI from the WIYN
3.5-m telescope in $B$, $V$, and $I$.\@  A deep color--magnitude
diagram is being constructed which will reveal the distance and
stellar populations characteristics of And VI.\@  We will report all our
findings on And VI in an upcoming journal paper.

\section{Discussion}
The discovery of the M31 dwarf spheroidal And V and the probable M31
dwarf spheroidal And VI increases the number of M31 dwarf spheroidals
from three to certainly four, and probably five.  It changes somewhat
the properties of M31's satellite system, as discussed below.  The most
obvious change is that M31 is not as poor in dwarf spheroidals as
previously thought.

Karachentsev (1996) discussed the spatial distribution of the
companions to M31.  The discovery of And V changes somewhat the spatial
distribution of the M31 satellites.  Curiously, And I, II \& III are
all located south of M31, while the three more luminous dwarf
elliptical companions NGC 147, 185 \& 205 are all positioned north of
M31.\@  Also, Karachentsev (1996) noted that there are more M31
companions overall south of M31 than north of M31.\@  And V's location
north of M31 lessens both of these asymmetries.  If And VI proves to be
a bona fide M31 companion, its location south of M31 would contribute
to the asymmetry.  With a projected radius from the center of M31 of
112 kpc, And V increases slightly the mean projected radius of the M31
dwarf spheroidals from 86 kpc to 93 kpc.  If And VI is also an M31
dwarf spheroidal, with its projected distance from M31 of 271 kpc, the
mean projected radius of the M31 dwarf spheroidals increases to 128
kpc.

The discovery of nearby dwarf galaxies like And V augments the faint
end of the luminosity function of the Local Group.  We do not yet have
a reliable $M_V$ value for And V, but it appears to be similar to that
of And III ($M_V$ = --10.2) since they have similar
extinction-corrected central surface brightness.  From a survey of nine
clusters of galaxies, Trentham (1998) derived a composite luminosity
function that is steeper at the faint end than that of the Local Group
(see his Figure 2).\@  He attributed the difference to poor counting
statistics and/or incompleteness among the Local Group sample.  The
discovery of And V reduces slightly the discrepancy between the Local
Group luminosity function and the extrapolation of Trentham's (1998)
function.  If And VI is a bona fide M31 dwarf spheroidal, it also
reduces this discrepancy.

It is also of interest to learn whether dwarf spheroidals as faint as
Draco and Ursa Minor ($M_V \approx -9$) exist around M31.\@  By
completing our survey and by understanding our completeness limits, we
plan to address this question.


\begin{references}
\reference{} Armandroff, T.\ E.\ 1994, in ESO/OHP Workshop on Dwarf
Galaxies, ed.\ G.\ Meylan \& P.\ Prugniel (Garching: ESO), 211
\reference{} Armandroff, T.\ E., Da Costa, G.\ S., Caldwell, N., \&
Seitzer, P.\ 1993, \aj, 106, 986
\reference{} Armandroff, T.\ E., Davies, J.\ E., \& Jacoby,
G.\ H.\ 1998, \aj, 116, 2287
\reference{} Brinks, E., \& Grebel, E.\ 1998, Dwarf Tales Newsletter,
Number 3
\reference{} Caldwell, N., Armandroff, T.\ E., Seitzer, P., \& Da
Costa, G.\ S.\ 1992, \aj, 103, 840
\reference{} Da Costa, G.\ S., \& Armandroff, T.\ E.\ 1990, \aj, 100,
162
\reference{} Da Costa, G.\ S., Armandroff, T.\ E., Caldwell, N., \&
Seitzer, P.\ 1996, \aj, 112, 2576
\reference{} Irwin, M.\ J.\ 1994, in ESO/OHP Workshop on Dwarf
Galaxies, ed.\ G.\ Meylan \& P.\ Prugniel (Garching: ESO), 27
\reference{} Karachentsev, I.\ 1996, \aap, 305, 33
\reference{} Konig, C.\ H.\ B., Nemec, J.\ M., Mould, J.\ R., \&
Fahlman, G.\ G.\ 1993, \aj, 106, 1819
\reference{} Lasker, B.\ M., \& Postman, M.\ 1993, in ASP
Conf.\ Ser.\ 43, Sky Surveys:  Protostars to Protogalaxies,
ed.\ B.\ T.\ Soifer (San Francisco: ASP), 131
\reference{} Lee, M.\ G., Freedman, W.\ L., \& Madore, B.\ F.\ 1993,
\apj, 417, 553
\reference{} Lee, Y.-W., Demarque, P., \& Zinn, R.\ 1990, \apj, 350,
155
\reference{} Mould, J., \& Kristian, J.\ 1990, \apj, 354, 438
\reference{} Reid, I.\ N., Brewer, C., Brucato, R.\ J., McKinley,
W.\ R., Maury, A., Mendenhall, D., Mould, J.\ R., Mueller, J.,
Neugebauer, G., Phinney, J., Sargent, W.\ L. W., Schombert, J., \&
Thicksten, R.\ 1991, \pasp, 103, 661
\reference{} Renzini, A., \& Buzzoni, A.\ 1986, in Spectral Evolution
of Galaxies, ed.\ C.\ Chiosi \& A.\ Renzini (Dordrecht: Reidel), 195
\reference{} Trentham, N.\ 1998, \mnras, 294, 193
\reference{} van den Bergh, S.\ 1972a, \apj, 171, L31
\reference{} van den Bergh, S.\ 1972b, \apj, 178, L99
\reference{} van den Bergh, S.\ 1974, \apj, 191, 271
\reference{} Whiting, A.\ B., Irwin, M.\ J., \& Hau, G.\ T.\ 1997, \aj,
114, 996
\end{references}
\end{document}